\documentstyle[amssymb]{article}

\newcommand{\bc}{\begin{C}}
\newcommand{\ec}{\end{C}}
\newcommand{\be}{\begin{equation}}
\newcommand{\ee}{\end{equation}}

\newcommand{\claim}{\begin{Cl}}
\newcommand{\eclaim}{\end{Cl}}
\newcommand{\nb}{\begin{Nb}}
\newcommand{\nbe}{\end{Nb}}
\newcommand{\bl}{\begin{LE}}
\newcommand{\el}{\end{LE}}

\newtheorem{Cl}{Claim}

\newcommand{\bd}{\begin{Def}}
\newcommand{\ed}{\end{Def}}
\newcommand{\bt}{\begin{Th}}
\newcommand{\et}{\end{Th}}

\newtheorem{Th}{Theorem}
\newtheorem{LE}{Lemma}
\newtheorem{C}{Corollary}
\newtheorem{Nb}{Note}
\newtheorem{Def}{Definition}

\begin{document}
 \title{Linear logic with idempotent exponential modalities: a note
}
\author{Sergey Slavnov
\\  National Research University Higher School of Economics, Moscow
\\ sslavnov@yandex.ru\\} \maketitle

\begin{abstract}

 In this note we discuss a variant of linear logic with idempotent exponential modalities. We propose a sequent calculus system and discuss its semantics. We also give a concrete relational model for this calculus.
\end{abstract}

\section{Introduction}
It is well known that the exponential modality in linear logic is ``not canonical'', in the sense that it is not uniquely determined by the logic rules. In concrete words, we can introduce several different copies of exponential connectives, obeying the same rules, but nothing implies that the connectives are equivalent. On the semantic side, this means that a model of multiplicative-additive linear logic (i.e. a $*$-autonomous category with products) may possess several non-equivalent structures, modeling the exponential fragment.

In general, we may say that the exponential fragment is  understood somewhat worse, at least on the semantic side, than the multiplicative-additive linear logic. Indeed, we know so many  models of ${\bf MALL}$ and ${\bf MLL}$; we know  concrete and abstract models, some complete  ones, some intuitive ones. We also know some abstract construction for producing these models. On the other hand, models of exponentials are rare, sometimes incorrect (this is often the case for interpretations based on the linear algebra and functional analysis setting), and arguably we have little understanding of how do these models arise.

Finally, some alternative exponentials are considered in literature, for example in bounded linear logic.

These lengthy remarks were to give some motivation for the topic of this paper, that is, to try convincing the reader that studying some non-standard exponentials might be of interest.
Specifically, we are considering linear logic with {\it idempotent} exponentials. This idea came to the author mainly from semantic considerations of relational models; it seems that in this setting a possibility of such an idempotent version is suggested rather naturally. On the other hand, highly informal and partial ``quantum-mechanical'' interpretations of ${\bf LL}$, discussed sometimes in literature, also may suggest idempotent exponentials; exponential modality may be tentatively understood as denoting some sort of a ``classical limit'' or a ``classicality property''. This point of view is taken in the work of Peter Selinger and Beno\^it Valiron on {\it quantum lambda calculus} {\cite{Selinger}}, where the typing system is based on linear logic.

Despite these considerations, as far as standard linear logic is concerned, models with idempotent exponentials are extremely degenerate and seem rather marginal. It may be interesting, thus, to search for a non-standard variant of ${\bf LL}$, where this idempotency is required by the structure; a version of linear logic where iterated exponentials are equivalent. This is the matter we deal with in this paper. We propose the system of {\it idempotent linear logic} (${\bf IdLL}$) and discuss its semantics.

 A subtle point, here, is that, {\it on the level of provability},  already the standard ${\bf LL}$ proves that the formulas $!A$ and $!!A$ are equivalent. Accordingly, our system of ${\bf IdLL}$  coincides with ${\bf LL}$ on this level;  it has the same set of provable formulas.  The difference between the two systems is {\it on the level of proofs}. This feature seems to us rather amusing.

\section{Linear Logic, comonads etc}
We assume that the reader is familiar with linear logic (${\bf LL}$) as well as with its categorical interpretation. For an introduction to the subject see, for example, \cite{Girard} and \cite{Mellies}.

In order to fix the terminology we recall that, semantically, linear logic describes a {\it $*$-autonomous category}, the $*$-autonomous structure being given by multiplicative connectives and the operations of linear negation and linear implication. The $*$-autonomous structure corresponding to the full linear logic carries also {\it products} and {\it coproducts}, given by the additive connectives.

In the following we  agree that a $*$-autonomous structure on the category ${\bf C}$ is specified by the {\it monoidal (``tensor'') product} $\otimes$, with the {\it monoidal unit} ${\bf 1}$, the {\it internal homs} functor $\multimap$, and the {\it dualizing object} $\bot$. Other relevant constructions, such as duality $(.)^*$, we understand as derived. In particular, $A^*=A\multimap \bot$. (We choose a star to denote duality in the models, because it seems to us more consistent with the general mathematical practice. However, for duality of ${\bf LL}$ formulas, we preserve the traditional notation $(.)^{\bot}$.)

To keep with linear logic notation, we denote product on a $*$-autonomous category as $\&$ and the corresponding neutral object as $\top$. The coproduct structure, then, is derived from duality.

As for the exponential fragment, which  interests us most in the current paper, its categorical formalization has some variations, which have been  discussed in literature; see \cite{Mellies} for a survey. Basically,  the $!$-modality corresponds to a {\it monoidal comonad}, that is to say a comonad $!$ satisfying
\be\label{monoidal-comonad}
!(A\& B)\cong !A\otimes !B,\mbox{   }
!\top\cong {\bf 1}.
\ee
(See \cite{MacLane} for a text-book definition and discussion of comonads.)

We  take a popular view that linear logic $!$-modality is a monoidal comonad coming from the following construction.

Let ${\bf C}=({\bf C},\otimes, {\bf 1}, \multimap, \bot, \&,\top)$ be a $*$-autonomous category with products, and ${\bf K}=({\bf K},\times, *)$ be a category with products and the neutral object $*$ . Let $F:{\bf C}\to{\bf K}$ and $G:{\bf K}\to{\bf C}$ be adjoint functors with the property that $F(\top)\cong\{*\}$, $F(A\& B)\cong F(A)\times F(B)$ and $G(*)\cong {\bf 1}$, $G(A\times B)\cong G(A)\otimes G(B)$. Then the composite functor $G\circ F:{\bf C}\to{\bf K}$ is a monoidal comonad on ${\bf C}$. The exponential $!$-modality is usually interpreted as a comonad of this form. 

(That the composition of two adjoint functors is a comonad is well known; see \cite{MacLane}, 6.1.
Since the functor $F$  preserves products, and the functor $G$ takes products to tensor products, it follows that this comonad is monoidal.)
Constructions of this sort  are known in the current literature on the subject as {\it linear-nonlinear models}, these were introduced in \cite{Benton}.

We are going to discuss the special setting when the above monoidal comonad is {\it idempotent}. That is, we have a $*$-autonomous category ${\bf C}=({\bf C},\otimes, {\bf 1}, \multimap, \bot)$ with products $(\&,\top)$ and a monoidal comonad $!:{\bf C}\to{\bf C}$ as above, with the corresponding comonadic natural transformations $\delta: !\to !!,\epsilon:!\to Id$, such that the natural transformation $\delta: !\to !!$ is an {\it isomorphism}.

As far as linear logic is concerned, this setting is, indeed, special. Usually, models of ${\bf LL}$ discussed in literature are {\it not} idempotent. And, of course,  the system ${\bf LL}$ itself, seen as a category, does not belong to this setting. However, as we are trying to show below, this special structure can be captured in a self-consistent proof-system, which we call {\it idempotent linear logic} (${\bf IdLL}$). It turns out also that this system has a very simple concrete model in the setting of {\it totality spaces}.  We think therefore that the structure of idempotent comonad on a $*$-autonomous category might be of some interest.

\section{Sequent calculus ${\bf IdLL}$}
The language of idempotent linear logic ($\bf IdLL$) coincides with that of linear logic. Formulas  are built from {\it positive} and {\it negative} literals, respectively,  $p_0, \ldots,p_n,\ldots$ and $p_0^\bot,\ldots,p_n^\bot,\ldots$, by means of the multiplicative and additive connectives $\otimes$, $\wp$, $\&$, $\oplus$ and the exponential modalities $!$ and $?$.  Linear negation $A^\bot$ of the formula $A$ is defined inductively by
$$(p^\bot)^\bot=p,\mbox{ } (X\otimes Y)^\bot=X^\bot\wp Y^\bot,\mbox{ }(X\wp Y)^\bot=X^\bot\otimes Y^\bot,$$
$$(X\& Y)^\bot=X^\bot\oplus Y^\bot,\mbox{ }(X\oplus Y)^\bot=X^\bot\& Y^\bot,$$
$$(!A)^\bot=?A^\bot,\mbox{ }(?A)^\bot=!A^\bot.$$
Linear implication is  defined by
$$
A\multimap B=A^\bot\wp B.
$$
Notation $!^nA$ and $?^nA$, as usual, stands for iterated modalities, that is for the formula $A$ preceded by $n$ $!$'s or $?$'s.

The sequent calculus for ${\bf IdLL}$ coincides with that of ${\bf LL}$ on the level of multiplicative-additive connectives:
 $$\frac{}{\vdash A,A^\bot}(Identity),\mbox{ }\frac{\vdash\Gamma, A\mbox{ }\vdash A^\bot, \Delta }{\Gamma\vdash \Delta} (Cut),$$
 $$\frac{\vdash\Gamma,A,B,\Delta}{\vdash\Gamma,B,A,\Delta}(Exchange), $$
 $$\frac{\vdash \Gamma,A\mbox{ }\vdash B,\Delta}{\vdash\Gamma,A\otimes B,\Delta}(Times),\mbox{ }\frac{\vdash \Gamma,A,B,\Delta}{\vdash\Gamma,A\wp B,\Delta}(Par), $$
$$\frac{\vdash \Gamma,A\mbox{ }\vdash\Gamma, B}{\vdash\Gamma,A\& B,\Delta}(With),\mbox{ }\frac{\vdash \Gamma,A}{\vdash\Gamma,A\oplus B}\mbox{ or }\frac{\vdash \Gamma,B}{\vdash\Gamma,A\oplus B}(Plus), $$
but somewhat differs for the exponential fragment. The rules are:
$$\frac{\vdash\Gamma,?A,?A}{\vdash\Gamma,?A} (Contraction),\mbox{ }\frac{\vdash\Gamma}{\vdash\Gamma,?A}(Weakening),$$ and
$$\frac{\vdash\Gamma,A}{\vdash\Gamma,?^nA}\mbox{ if the main connective of }A\mbox{ is not }?\mbox{ }(n-Dereliction),$$
$$\frac{\vdash?A_1,\ldots,?A_k, A}{\vdash ?A_1,\ldots,?A_k, !^nA}\mbox{ if the main connective of }A\mbox{ is not }!\mbox{ }(n-Promotion).$$

Thus ${\bf IdLL}$ differs from ${\bf LL}$ only in the rules of Dereliction and Promotion for introducing exponential connectives. Recall that for ${\bf LL}$ we have the rules
$$\frac{\vdash\Gamma,A}{\vdash\Gamma,?A}(Dereliction),$$
$$\frac{\vdash?A_1,\ldots,?A_k, A}{\vdash ?A_1,\ldots,?A_k, !A}(Promotion).$$

Furthermore, it can be seen very easily that, on the level of {\it provability}, the two systems simply coincide, i.e. ${\bf IdLL}$ and ${\bf LL}$ have the same sets of provable sequents.

Indeed, $n$-Promotion and $n$-Dereliction are admissible in ${\bf LL}$ as iterations of Promotion and Dereliction, respectively. On the other hand, ${\bf IdLL}$ obviously derives the sequents $\vdash ?^nA,!^mA^{\bot}$ for all $n,m>0$, and, with the use of Cut, this allows us to emulate Promotion and Dereliction in this system. Typically, if $A=?^nA'$ is a formula, having $?$ as its main connective, where $A'$ does not start with $?$, and we have  an ${\bf IdLL}$-derivation  of $\vdash \Gamma, A$,  i.e. of $\vdash \Gamma, ?^nA'$, then by cutting this sequent with $\vdash !^n(A')^\bot,?^{n+1}A'$ we derive $\vdash \Gamma, ?^{n+1}A'$, i.e. $\vdash \Gamma, ?A$, as if we had Promotion. The case of Dereliction is treated identically. Let us write down this simple conclusion as a theorem.

\bt The systems ${\bf LL}$ and ${\bf IdLL}$ have the same sets of provable sequents.
\et
\bigskip

The next observation is that ${\bf IdLL}$ is cut-free. Essentially, cut-elimination algorithm for ${\bf IdLL}$ is the same as for ${\bf LL}$. When the cut formulas are of the form $A=?^nA'$, $A^\bot=!^n(A')^\bot$, with $A'$ not starting with $?$, we should treat the initial segment of $?$'s ($!$'s) as a single main connective.
\bt ${\bf IdLL}$ enjoys cut-elimination.
\et
{\bf Proof} The cut-elimination algorithm and its correctness proof can be taken verbatim from \cite{CutElim}, with words {\it Dereliction} and {\it Promotion} replaced with {\it n-Dereliction} and {\it n-Promotion}, respectively, and, accordingly, $?$ and $!$ replaced with $?^n$ and $!^n$.
\bigskip

Now, having established the cut-elimination property, we can compare ${\bf IdLL}$ with ${\bf LL}$ not only on the level of provability, but on the level of proofs. In fact it is easy to see that ${\bf IdLL}$ has strictly fewer cut-free proofs: for the literal $p$ there is only one way to derive $\vdash ?^np^\bot,!^np$ without Cut in ${\bf IdLL}$ ($n$-Dereliction followed by $n$-Promotion), which is not at all the case for ${\bf LL}$.

Concretely, now we can say in some precise sense that ${\bf IdLL}$ treats formulas $??A$ and $?A$ as {\it isomorphic}, and similarly with $!$. (This justifies the title {\it idempotent}.)

Let $\pi_1$ and $\pi_2$ be ${\bf IdLL}$ proofs of the sequents $\vdash??A,!A^\bot$ and $\vdash ?A,!!A^\bot$; the first proof is obtained from the Identity axiom by 2-Dereliction followed by 1-Promotion, the second one, by 1-Dereliction followed by 2-Promotion. We can put the two proofs together and then apply the Cut-rule to their conclusions in two ways: the pair of cut-formulas being either $!!A^\bot/??A$ or $!A^\bot/?A$. Thus we obtain a proof of $\vdash?A,!A^\bot$ and a proof of $\vdash ??A,!!A^\bot$. However it is immediate that, after cut-elimination, both proofs normalize to corresponding Identity axioms. In other words,  from the categorical point of view, when proofs are seen as morphisms, it turns out that the above proofs $\pi_1$ and $\pi_2$ are mutually inverse and establish an isomorphism $?A\cong ??A$. Thus we can claim that the system ${\bf IdLL}$ gives an accurate sequent calculus axiomatization for the above defined structure of $*$-autonomous category with an idempotent comonad.

\bt The category of  formulas and cut-free   ${\bf IdLL}$ proofs is $*$-autonomous with an idempotent comonad induced by the $!$-connective.
\et

\section{Idempotent comonad on totality spaces}
In this Section we  discuss Loader's {\it totality spaces} (see \cite{Loader}), a very well known and, in some sense, very natural model of linear logic, similar to Girard's coherence spaces.  It turns out that totality spaces provide a concrete model of idempotent comonadic structure.

We use the following definitions.
A {\it pre-totality space} $A$ is a pair $A=(|A|,A_{tot})$, where $|A|$ ({\it base} of $A$) is a set, and $A_{tot}\subseteq 2^{|A|}$. The elements of $A_{tot}$ are called {\it total} sets of $A$.

 The {\it dual} $A^*$ of $A$ is the pre-totality space $A^*=(|A|,A^*_{tot})$, where $A^*_{tot}\subseteq 2^{|A|}$ consists of all subsets $r\subseteq|A|$, satisfying the condition: $\forall s\in A_{tot}$ the set $r\cap s$ is a singleton.

 A {\it totality space} $A$ as a pre-totality space coinciding with its bidual (i.e., having the same total sets), $A=A^{**}$. A standard observation is that the dual of a pre-totality space is a totality space. The total sets for $A^*$ are called {\it cototal} for $A$.

We define the following operations on totality spaces, corresponding to multiplicative and additive connectives of ${\bf LL}$ (i.e. to the $*$-autonomous structure).

The {\it tensor product} $A\otimes B$ of two totality spaces $A$ and $B$ has the set $|A|\times|B|$ as its base, and total sets of the form $r\times s$, where $r\in A_{tot}$, $s\in B_{tot}$. It can be shown (see \cite{Loader}) that the totality space $A\otimes B$ is well-defined, i.e $A\otimes B=(A\otimes B)^{**}$. Note that
\be\label{tensor}
(A\otimes B)_{tot}\cong A_{tot}\times B_{tot}.
\ee

The {\it cotensor product} $A\otimes B$ of two totality spaces $A$ and $B$ is defined simply as the dual $A\wp B=(A^*\otimes B^*)^*$.

The additive operations (product and coproduct) of $A$ and $B$ have the disjoint union of
$|A|$ and $|B|$ as their bases. Total sets of $A\& B$ are disjoint unions of total sets of $A$ and $B$, whereas total sets of $A\oplus B$ are total sets of $A$ and total sets of $B$. It is easy to see that the two operations are dual: $(A\& B)^*=A^*\oplus B^*$ and $(A\oplus B)^*=A^*\& B^*$.  Note also that
\be\label{prod}
(A\& B)_{tot}\cong A_{tot}\times B_{tot}.
\ee

The category ${\bf Tot}$ has totality spaces as its objects, and total sets of $A\multimap B=A^*\wp B$ as morphisms between $A$ and $B$, the composition being that of relations. The category ${\bf Tot}$ is known to be $*$-autonomous (with monoidal units ${\bf 1}=\bot=(\{*\},\{\{*\}\})$).

Finally let us define {idempotent exponentials} on ${\bf Tot}$.

\bd For a totality space $A$ let $!A$ be the totality space with the base $A_{tot}$, whose total sets are all singletons.
\ed

It is straightforward that the dual $(!A)^*$ is the pair $((A_{tot}, \{A_{tot}\}))$, and $!A=!A^{**}$, so this is indeed a totality space.

Note that it immediately follows from (\ref{tensor}) and (\ref{prod}) that
\be
!(A\& B)\cong !A\otimes !B.
\ee

The operation is clearly idempotent: $!A\cong !!A$. Let us show that it is indeed a comonad.

We have the (cartesian closed) category ${\bf Sets}$ of sets and functions. To every set $S$ we can associate a {\it discrete} totality space $Dis(S)$ with base $S$ and all singletons as total sets. For any function $f: S\to T$, its graph $grapf(f)$ is a relation in $S\times T$. Moreover, for any $x\in S$, this relation has exactly one point in the intersection with the set $\{x\}\times T$ (this is a fancy way to say that $f$ is total and single-valued). Since the above $\{x\}\times T$ is the most general form of a total set in $Dis(S)\otimes (Dis(T))^*$,  it follows that $graph(f)$ is total in $(Dis(S)^*)\wp Dis(T)=Dis(S)\multimap Dis(T)$, i.e. a morphism between corresponding objects in ${\bf Tot}$. In other words, we have the functor $Dis:{\bf Sets}\to{\bf Tot}$ (since composition, obviously, is preserved). Note also that the functor preserves monoidal structure: $Dis(S\times T)=Dis(S)\otimes Dis(T)$, $Dis(\{*\})={\bf 1}$.

We have another functor in the opposite direction. It associates to each object $A$ of ${\bf Tot}$ the set of total sets of $A$, and to each morphism $\phi$ between totality spaces $A$ and $B$ the obvious function between $A_{tot}$ and $B_{tot}$, induced by composition with $\phi$. (In fact, this functor is just the Yoneda embedding ${\bf{Tot}}({\bf 1},.)$.) Let us call this functor ${\bf Y}$. Note that the functor preserves the product structure: ${\bf Y}(A\& B)={\bf Y}(A)\times {\bf Y}(B)$, ${\bf Y}({\bf 1})\cong (\{*\})$.

Now we claim that the functors $Dis$ and ${\bf Y}$ are adjoint. Since our exponential $!$ is just their composition, this implies that it is a comonad.

Indeed, let $A$  be a totality space and $S$ be a set. Any morphism $\phi: Dis(S)\to A$ in ${\bf Tot}$ determines a (total, single-valued) function from  $Dis(S)_{tot}$ to $A_{tot}$. However, total sets of $Dis(S)$ are just elements of $S$, i.e. $Dis(S)_{tot}$ naturally identifies with $S$, whereas the set $A_{tot}$ is ${\bf Y(A)}$. Thus the ${\bf Tot}$-morphism $\phi$ determines a function, i.e. a ${\bf Sets}$-morphism $S\to{\bf Y(A)}$, and we have the map ${\bf Tot}( Dis(S),A)\to {\bf Sets}(S,{\bf Y}(A))$.

On the other hand, for a function $f:S\to A_{tot}$, consider the relation $\hat f=\{(s,a)|s\in S, a\in |A|,\mbox{ s.t. }a\in f(s)\}$. This relation is total in $Dis(S)\multimap A$. Indeed,  any cototal set $\alpha$ in $Dis(S)\multimap A$ is of the form $\alpha=\{s\}\times \tau$, where $s\in S$ and $\tau\subseteq |A|$ is cototal in $A$. Then the intersection  $\hat f\cap\alpha =\{s\}\times (f(s)\cap\tau)$ is a singleton  because $f(s)$ is total in $A$. Thus $\hat f$ is a morphism in ${\bf Tot}$, and we have the map ${\bf Sets}(S,{\bf Y}(A))\to{\bf Tot}( Dis(S),A)$. It is immediate that this map is inverse to the one defined in the preceding paragraph.

Thus, the functors ${\bf Y}$ and $Dis$ are adjoint, hence $!=Dis\circ{\bf Y}$ is a comonad. Moreover, the functor ${\bf Y}$ sends product in ${\bf Tot}$ to product in ${\bf Sets}$, and $Dis$ sends product in ${\bf Sets}$ to tensor product in ${\bf Tot}$. Thus $!$ sends product in ${\bf Tot}$ to tensor product, hence it equips each object $!A$ with the comonoid structure with respect to tensor product induced by the comonoid structure of $A$ with respect to the product. In other words,  we can conclude with the theorem:
\bt The functor $!$ is an idempotent comonoidal comonad on the category ${\bf Tot}$.
\et

 \end{document}